\documentclass[]{aa}

\usepackage[varg]{txfonts}
\usepackage{color}
\usepackage{graphicx}
\usepackage{lscape}
\usepackage{longtable}
\usepackage{natbib}
\usepackage{ulem}
\usepackage[utf8]{inputenc}
\usepackage{graphicx}
\usepackage{epstopdf}
\usepackage{amsmath}
\bibpunct{(}{)}{;}{a}{}{,} 
\usepackage[switch]{lineno}

\begin{document}

\title{Observation of quasi-periodic solar radio bursts associated with propagating fast-mode waves}
\author{C. R. Goddard \inst{1} \and G. Nistic\`o \inst{1} \and V. M. Nakariakov \inst{1,2,3} \and I. V. Zimovets \inst{4,5,6} \and S. M. White \inst{7}}
\institute{Centre for Fusion, Space and Astrophysics, Department of Physics, University of Warwick, CV4 7AL, UK, \email{c.r.goddard@warwick.ac.uk}\
\and Astronomical Observatory at Pulkovo of the Russian Academy of Sciences, 196140 St Petersburg, Russia 
\and School of Space Research, Kyung Hee University, 446-701 Yongin, Gyeonggi, Korea
\and Space Research Institute (IKI) of Russian Academy of Sciences, Profsoyuznaya St. 84/32, 117997 Moscow, Russia 
\and  State Key Laboratory of Space Weather, National Space Science Center, Chinese Academy of Sciences, Zhongguancun Nanertiao 1, Haidian District, 100190 Beijing, China
\and International Space Science Institute, Zhongguancun Nanertiao 1, Haidian District, 100190 Beijing, China
\and Air Force Research Laboratories, Space Vehicles Directorate, Albuquerque, NM, 87117, USA
}

\date{Received 10/03/2016 /Accepted 08/08/2016}

\abstract 
{}
{Radio emission observations from the Learmonth and Bruny Island radio spectrographs are analysed to determine the nature of a train of discrete, periodic radio \lq sparks\rq (finite-bandwidth, short-duration isolated radio features) which precede a type II burst. We analyse extreme ultraviolet (EUV) imaging from SDO/AIA at multiple wavelengths and identify a series of quasi-periodic rapidly-propagating enhancements, which we interpret as a fast wave train, and link these to the detected radio features.}
{The speeds and positions of the periodic rapidly propagating fast waves and the coronal mass ejection (CME) were recorded using running-difference images and time-distance analysis. From the frequency of the radio sparks the local electron density at the emission location was estimated for each. Using an empirical model for the scaling of density in the corona, the calculated electron density was used to obtain the height above the surface at which the emission occurs, and the propagation velocity of the emission location.}
{The period of the radio sparks, $\delta t_\mathrm{r}$=1.78$\pm$0.04 min, matches the period of the fast wave train observed at 171 $\AA$, $\delta t_\mathrm{EUV}$=1.7 $\pm$ 0.2 min. The inferred speed of the emission location of the radio sparks, 630 km s$^{-1}$, is comparable to the measured speed of the CME leading edge, 500 km s$^{-1}$, and the speeds derived from the drifting of the type II lanes. The calculated height of the radio emission (obtained from the density) matches the observed location of the CME leading edge. From the above evidence we propose that the radio sparks are caused by the quasi-periodic fast waves, and the emission is generated as they catch up and interact with the leading edge of the CME.}
{}

\keywords{Sun: corona - waves - oscillations - radio radiation - type II bursts - EUV waves - flares - CMEs - methods: observational}
\maketitle

\titlerunning{Quasi-periodic radio emission associated with MHD waves}
\authorrunning{Goddard et al.}

\section{Introduction}

Flaring activity on the Sun has been studied since the first detections of solar flares. This study has intensified over the last two decades due to advancements in instrumentation and in response to the need for plasma astrophysics and space weather research. Flaring activity triggers waves and oscillations in the solar corona, the study of which allows comparisons to models and theory of magneto-hydrodynamics (MHD) waves to be made, as well as seismological inversions which allow the local plasma parameters to be measured indirectly \citep[see][]{2005LRSP....2....3N, 2012RSPTA.370.3193D, Nakariakov2016}. In the extreme ultraviolet (EUV) band detected wave activity has included large scale phenomena such as global \lq\lq EIT  waves\rq\rq \citep[e.g.][]{2009SoPh..259...49P, 2011SSRv..158..365G}, and their fast (coronal Moreton waves) and chromospheric counterparts \citep[e.g.][]{1960AJ.....65U.494M, 2005ApJ...622.1202C, 2015LRSP...12....3W}. Smaller scale activity includes decaying and decay-less kink oscillations of coronal loops \citep[e.g.][]{2015A&A...583A.136A, 2016A&A...585A.137G}, propagating and standing slow magnetoacoustic waves \citep[e.g.][]{2009SSRv..149...65D, 2011SSRv..158..397W} and rapidly propagating quasi-periodic wave trains \citep[e.g.][]{2011ApJ...736L..13L}.  

\cite{2011ApJ...736L..13L} detected EUV emission disturbances at 171 $\AA$, propagating from a flaring source along a coronal funnel, with a projected phase speed of 2000 km s$^{-1}$ and a 3 min period. \cite{2012ApJ...753...52L} detected wave trains running ahead of and behind a CME front at 171 and 193 $\AA$, with a dominant 2 min period. More recently, \cite{2014A&A...569A..12N} detected and modelled a fast coronal wave train propagating along two different paths, with a speed of $\leq$ 1000 km s$^{-1}$ and period of 1 minute. Similar detections include \cite{ 2011ApJ...736L..13L, 2012ApJ...753...53S} and \cite{2013A&A...554A.144Y}.

Signatures of waves and oscillations in the corona may be observed in the emission from solar flares. The observed quasi-periodic pulsations in flare emission (QPP's) \citep{1969ApJ...155L.117P} can be caused by MHD waves and oscillations \citep[e.g.][]{2010SoPh..267..329K}, although there are multiple other mechanisms for the quasi-periodic modulation of the emission \citep{2009SSRv..149..119N}. Recent observations show similar behaviour in multiple stellar flares \citep[e.g.][]{2016MNRAS.459.3659P}.

With the availability of imaging instruments simultaneously covering multiple wavelengths, and spatially resolved and unresolved recording of solar radio emission it becomes possible to study the relationship between MHD waves and oscillations and various non-thermal phenomena \cite[see, e.g.][]{2009A&A...505..791S}. One of the most intensively studied examples of this are type II radio bursts.

Coronal type II radio bursts are usually seen as two parallel emission lanes on solar radio spectrograms with an instant frequency ratio of approximately two, drifting from high to low frequencies. It is generally accepted that this radio emission is a result of plasma wave excitation at fronts of MHD shock waves propagating upwards through the corona. The lower and higher frequency lanes are thought to be emission at the fundamental and second harmonic of local plasma frequency, respectively \citep[e.g.][]{1966SvA.....9..572Z, 1995A&A...295..775M, 2008A&ARv..16....1P}.  The frequency drift of the lanes can be used to calculate the speed of the emission location, which is typically in the range of observed coronal mass ejection (CME) velocities. Despite this established association the physical relationship between flares, CMEs and the subsequent type II bursts is still only poorly understood.

The frequency of the emission from the MHD shock wave is given by the plasma frequency, 
\begin{align} F=8.98 \times 10^{-3} \sqrt{n_\mathrm{e}} MHz, \end{align} 
with the electron density, $n_\mathrm{e}$, in cm$^{-3}$. Empirical models for the scaling of the coronal density with height can be used to determine the height and speed of the emission location using the electron density obtained from the frequency. A commonly applied model is the Newkirk model \citep{1961ApJ...133..983N}, 
\begin{align}n_\mathrm{e}=n_\mathrm{e0} \times 10^{4.32(R_\mathrm{\odot}/R)},\end{align} 
where $n_\mathrm{e0}$=4.2$\times 10^{4}$ cm$^{-3}$. With this information it is possible to use EUV imaging observations to observe the emission location directly. Sometimes a splitting of the main emission lanes (fundamental and harmonic) of the type II bursts into two or more additional sub-lanes is observed. There is no consensus on this phenomenon yet, possibilities include simultaneous radio emission from the downstream and upstream regions of a shock \citep[e.g.][]{1974IAUS...57..389S, 2012A&A...547A...6Z}, multiple expanding structures in the CME creating multiple shocks, the passage of a single shock through surrounding structures  \citep[e.g.][]{1967PASAu...1...47M, 2012JGRA..117.4106S} or a combination of these \citep{2015AdSpR..56.2811Z}. 

Solar radio bursts can be modulated by variation in the density of the background plasma by MHD waves.  For example, Type IV bursts, broadband emission associated with high-energy non-thermal electrons accelerated during flares, exhibit fine structure and periodicity \citep[e.g.][]{2005ASSL..320..259M}. Analysis of light curves from radio and microwave wavelengths showed signatures of quasi-periodic wave trains \citep[e.g.][]{2009ApJ...697L.108M, 2011SoPh..273..393M}. Zebra patterns were found to show periodic wiggling, interpreted as the modulation of the double-plasma resonance location by magnetoacoustic sausage oscillations \citep{2013ApJ...777..159Y}.

In this paper we present the study of an individual flaring event, SOL2014-11-03T22:15, the subsequent CME and wave activity and the associated type II burst. We present compelling evidence that a series of quasi-periodic \lq sparks\rq\ in the radio spectra are linked to disturbances seen in the low corona in the EUV band. This is the first observation which links quasi-periodic fast waves observed in the EUV band to quasi-periodic features in radio spectra. In Sect. 2 the data and instruments are described, in Sect. 3  the observations and results are presented, and a discussion and summary are given in Sects 4 and 5.

\section{Instruments and data}

\begin{figure}
\resizebox{\hsize}{!}{\includegraphics{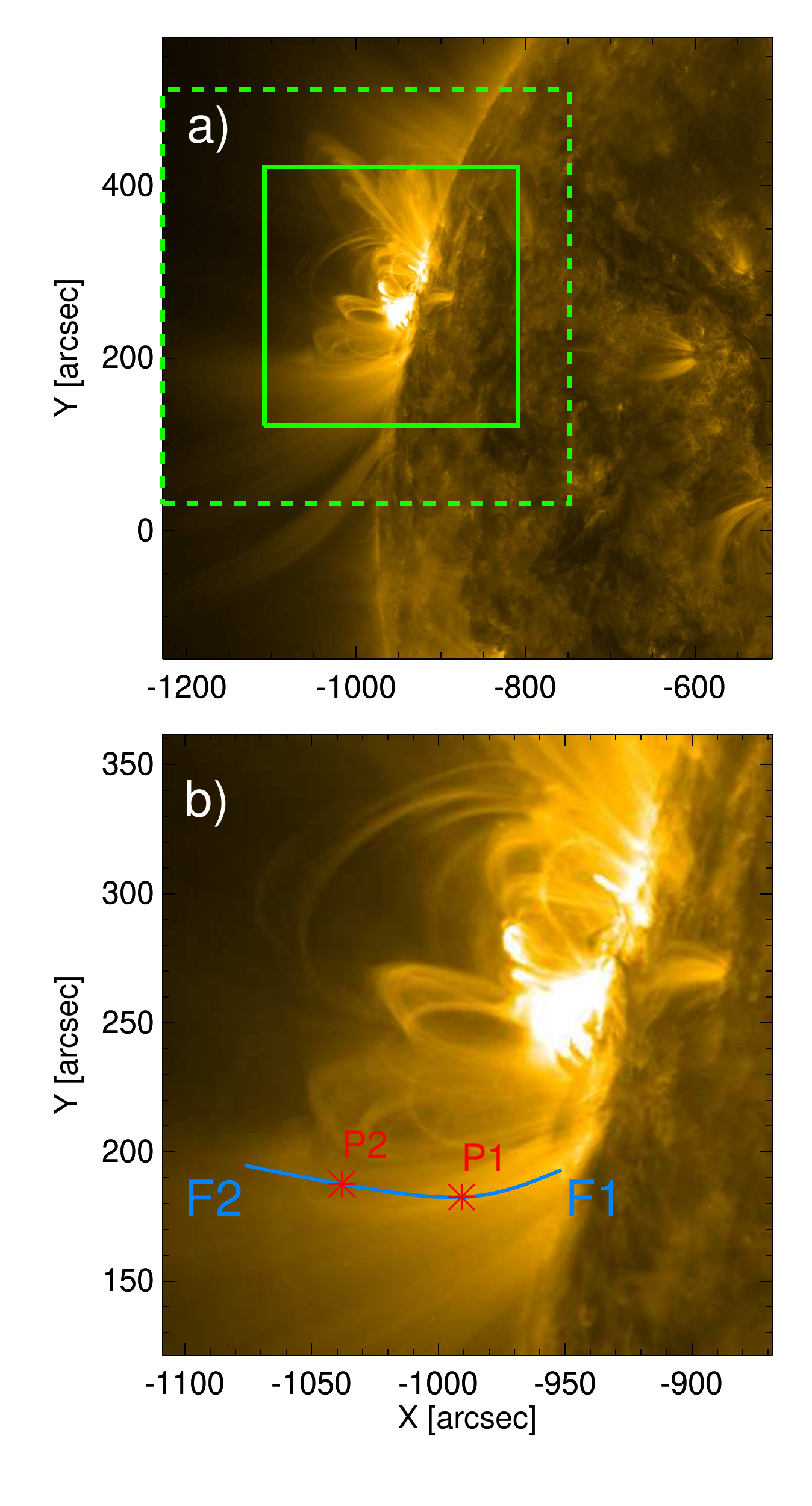}}
\caption{Panel a): SDO/AIA 171 $\AA$ image during the observed event at 22:00.00 UT centred on the active region of interest (AR 12205). The green boxes show the fields of view used for Fig. \ref{3_wt} (solid) and Fig. \ref{3_cme} (dashed). Panel b): the analysed region at 22:40:12 UT, F1 and F2 show the apparent extrema of the propagation path of the observed periodic intensity enhancements along a funnel structure, and the blue fit shows the slit used in the analysis. The red points P1 and P2 indicate the positions at which the time series plotted in Fig. \ref{td_maps} were extracted from.}
\label{ar_fig}
\end{figure}

\begin{figure*}
\resizebox{\hsize}{!}{\includegraphics{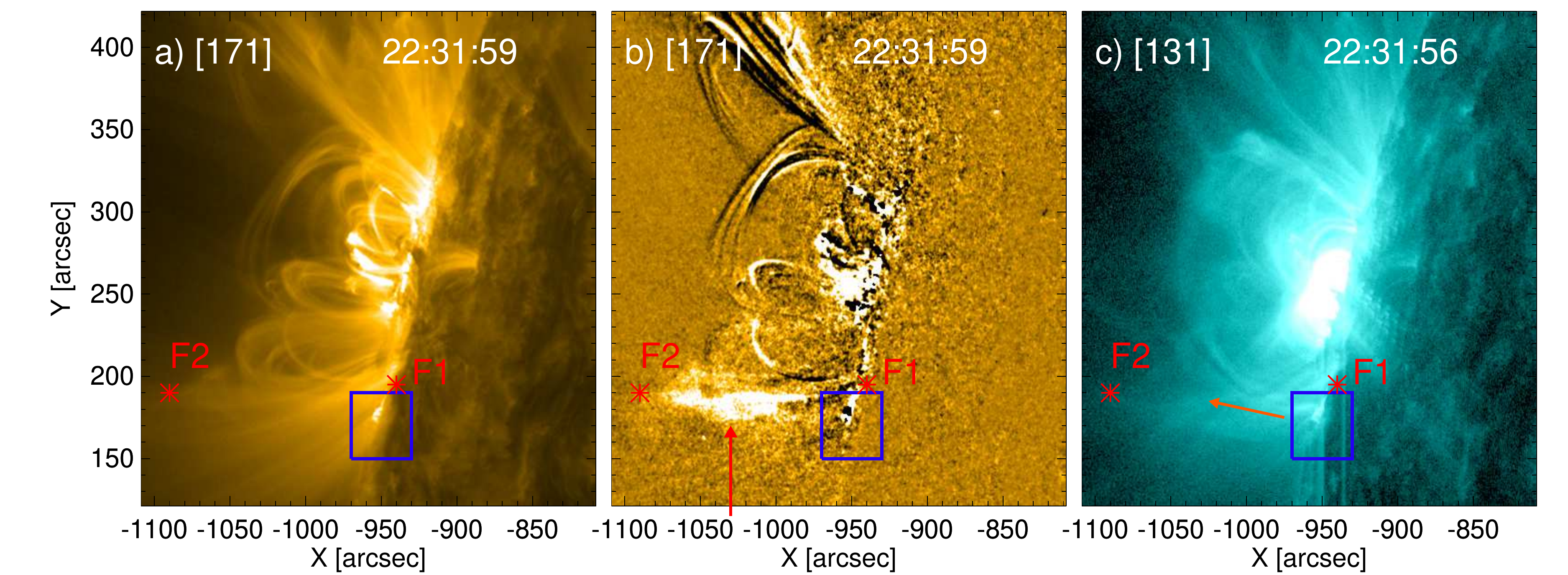}}
\caption{ Three SDO/AIA images of the active region. In all panels the region of enhanced emission, associated with the ejection of the EUV wave-train, is highlighted by a blue box. The red points F1 and F2 indicate the start and end points of a guiding funnel structure. Panel a) shows the 171 $\AA$ image. Panel b) also shows a 171 image, with the previous frame subtracted. The red arrow indicates an enhancement propagating along the funnel. Panel c) shows a 131 $\AA$ image, the orange arrow indicates the direction of a propagating outflow. The time evolution is shown in a movie available online.}
\label{3_wt}
\end{figure*}
 
	 EUV imaging was used from the Atmospheric Imaging Assembly \cite[AIA;][]{2012SoPh..275...17L} onboard the Solar Dynamics Observatory (SDO).  The EUV data sets were retrieved in the FITS format from the JSOC data centre\footnote{http://jsoc.stanford.edu/ajax/lookdata.html}, with spatial and temporal resolution of 0.6 arcsec per pixel and 12 s respectively, using the SSW function \texttt{vso\textunderscore search.pro}. The images were prepared and corrected using the standard SSW routine \texttt{aia\textunderscore prep.pro}, and normalised by the exposure time of the instrument, which varies during the flare emission. The cadence also varies by $\pm$ 1 s from 22:41:00 UT, so the data for subsequent frames was re-binned to a constant cadence when required. Data was downloaded between 22:00 UT and 23:30 UT, resulting in 450 frames of 4096 $\times$ 4096 pixels. Two smaller fields of view of 800 $\times$ 800 pixels (bottom left corner $x$=0, $y$=2100) and 500 $\times$ 500 pixels (bottom left corner $x$=200, $y$=2250) used in the processing and analysis are shown in panel a) of Fig. \ref{ar_fig}. 
	 
Radio spectra covering the range 25-180 MHz was obtained from Learmonth Solar Radio Observatory in Western Australia, part of the USAF Radio Solar Telescope Network (RSTN) \citep{learmonth...2003}. The data is arranged in two bands, 25-75 MHz and 75-180 MHz, and is linearly spaced in both. Supplementary data covering the range 6 to 62 MHz from the Bruny Island Radio Spectrograph (BIRS) \citep{1997PASA...14..278E}, located on Bruny Island off the south-eastern coast of Tasmania, was analysed to confirm the presence of features detected in Learmonth spectra qualitatively. 
	 
Additional data for the event could not be obtained, as the Hinode instruments were targeting a different region of the disk, and ground based radio instruments such as the Nan\c{c}ay Radioheliograph, the Nobeyama Radioheliograph and the Siberian Solar Radio Telescope missed the event due to their respective instrumental night times. No spatial information was available in the radio band, and RHESSI instruments recorded no data of interest during the event.



\section{Observations and analysis}

\begin{figure}
\resizebox{\hsize}{!}{\includegraphics{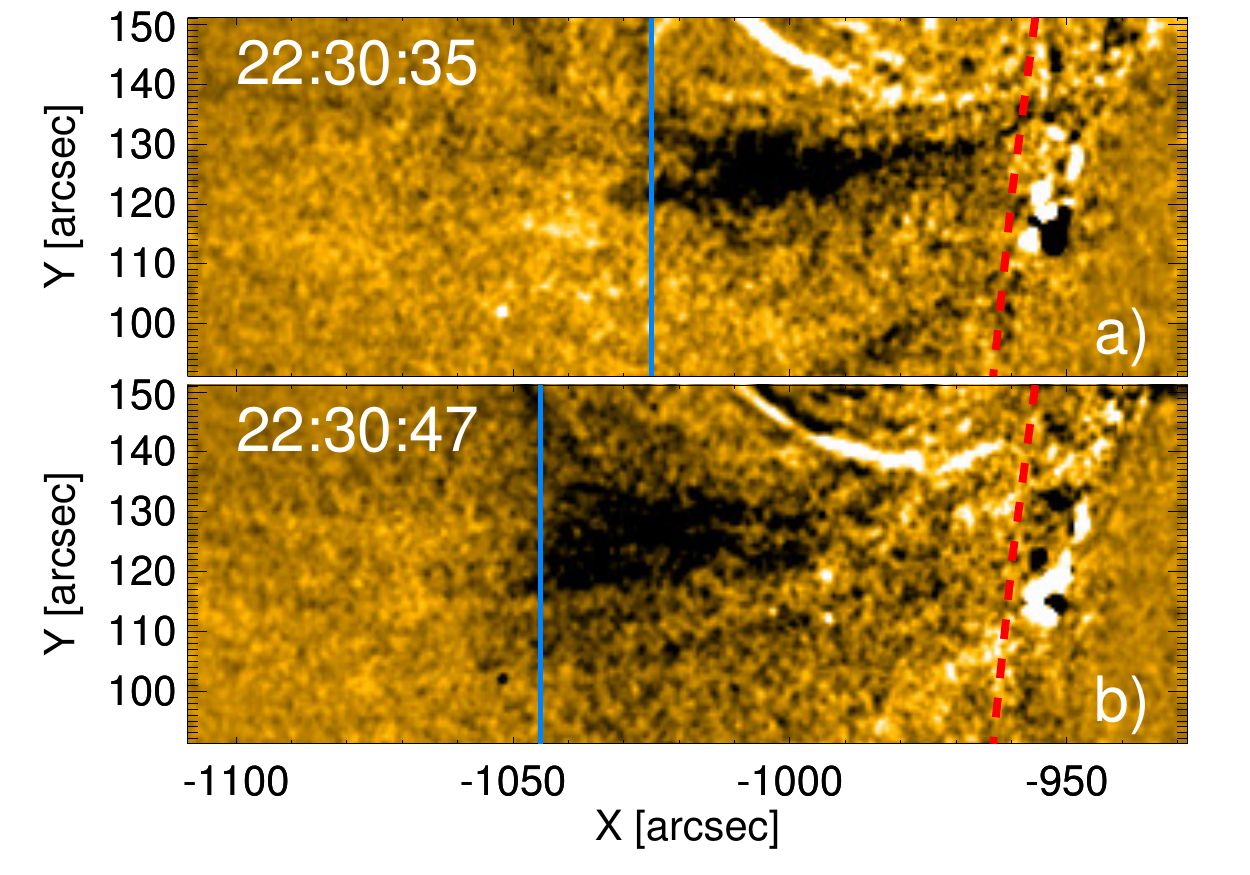}}
\caption{ Two running difference images of the funnel structure at 171 $\AA$. There is a one frame (12 seconds) separation between the two images. The vertical blue lines approximate the position of the propagating wave front. The dashed red curve indicates the position of the solar limb.}
\label{prop_fig}
\end{figure}

\begin{figure}
\resizebox{\hsize}{!}{\includegraphics{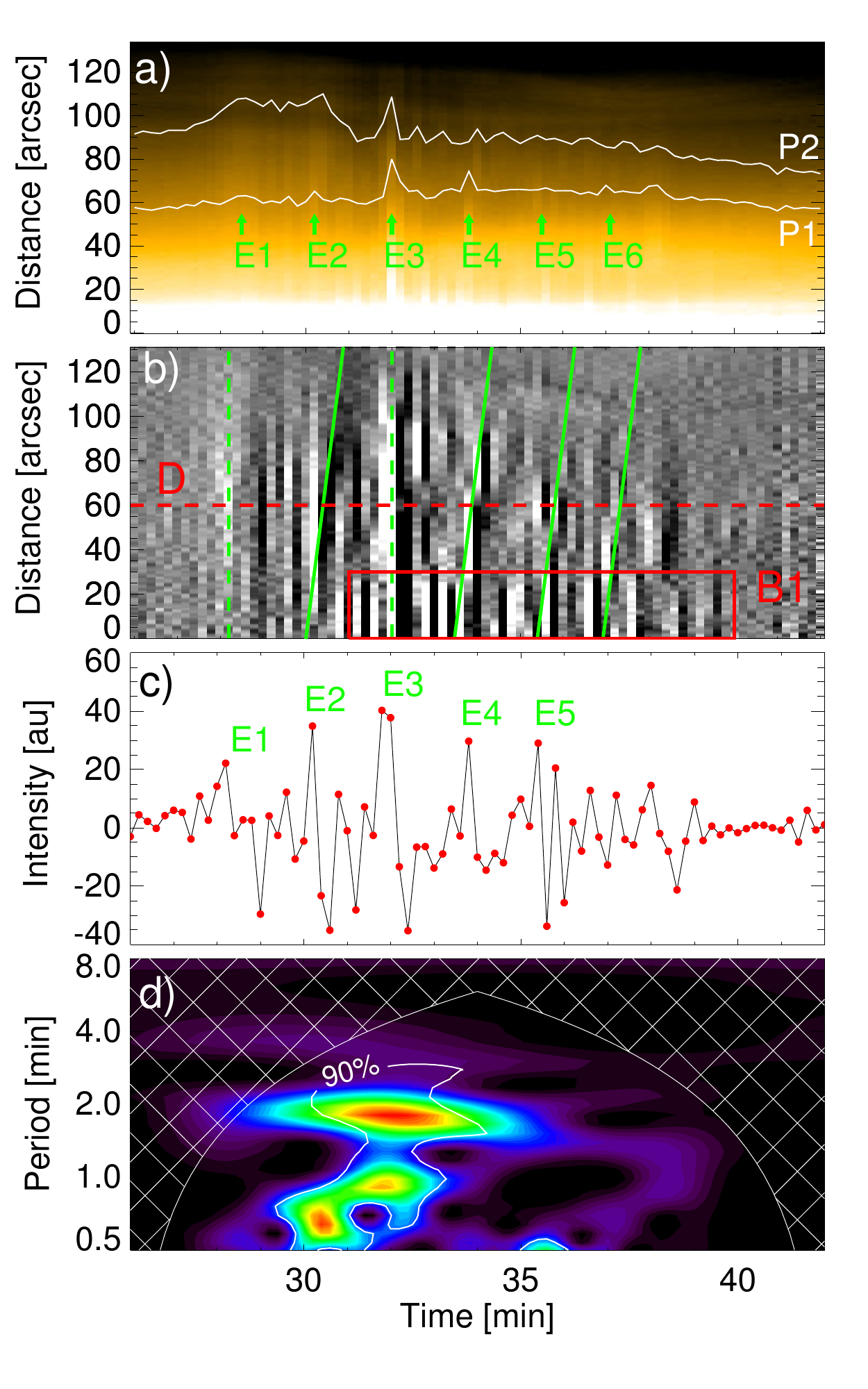}}
\caption{ \textit{Panel a)}: time distance (TD) map formed from the slit along the path of intensity enhancements, between points F1 and F2, shown in Fig. \ref{ar_fig}. The intensity profiles labelled P1 and P2 show the intensity profile at different positions along the slit, and the green arrows indicate the position of main peaks, labelled E1-E6. \textit{Panel b)}: a TD map formed from a slit along the path of intensity enhancements in the running difference images.  The diagonal green lines show the propagating intensity enhancements, and the dashed vertical lines indicate the enhancements where no propagation is seen. Box B1 highlights a series of shorter path, shorter period enhancements. \textit{Panel c)}: an intensity time series extracted from D in panel b), the intensity enhancements E1-E5 are labelled (E6 is missed as it is not prominent at the chosen distance along the slit). \textit{Panel d)}: Morlet wavelet spectra for the intensity time series, showing the distribution of the oscillation power with period as a function of time. The time axis of the four panels refers to the time elapsed since 22:00 UT. }
\label{td_maps}
\end{figure}

EUV and radio observations of a flaring event of GOES class M6.5 in the active region AR 12205 on the 3rd of November 2014 have been analysed. The active region is located on the eastern solar limb (see Fig. \ref{ar_fig}). The GOES X-ray lightcurves for the event, obtained using the SolarSoft (SSW) function \texttt{goes.pro}, show a characteristic peak at the time of the observations, beginning at approximately 22:06:30 UT, and reaching its peak at 22:39:30 UT. A coronal mass ejection was associated with the flare, with an average apparent speed of $\sim$ 500 km s$^{-1}$ according to the Computer Aided CME Tracking (CACTus) catalogue\footnote{http://sidc.oma.be/cactus/catalog.php} \citep{2009ApJ...691.1222R}. A global EUV wave was also triggered. 

\subsection{EUV observations}

A series of quasi-periodic intensity enhancements are seen within a guiding funnel structure in the 171 $\AA$ band, one is indicated by the red arrow in panel b) of Fig. \ref{3_wt}, between points F1 and F2. This series of enhancements will be referred to as \lq wave train\rq\ throughout the paper. This guiding structure is part of a bundle of open and expanding flux tubes, or funnels, to the south of the active region. This structure is similar to those analysed in \cite{2011ApJ...736L..13L} and \cite{2014A&A...569A..12N}, and modelled by \cite{2013A&A...560A..97P}, which were also found to guide propagating fast-mode wave trains.

The projected speed of the propagating wave train can be estimated from the observed distance the individual fronts move between frames. Using the positions indicated by the blue lines in Fig. \ref{prop_fig} and the time between the observations, 12 seconds, a speed of 1200 km s$^{-1}$ is obtained.

Time-distance (TD) maps are the standard method for visualising waves and oscillations in imaging data and determining their periods, amplitudes and phase speeds. A slit was created by selecting a series of points along the centre of the propagation path of interest (between extrema F1 and F2) and fitting them with a spline function, plotted in panel b) of Fig. \ref{ar_fig}. A TD map was formed by interpolating over the pixels crossed by the fit, and averaging over an 11 pixel width, giving a one-dimensional intensity profile for each frame which can be used to form an image, with one axis representing time and the other the distance along the slit. TD maps created from the normal intensity and running difference images, are shown in panels a) and b) of Fig. \ref{td_maps}. The intensity profiles at two different distances are plotted on panel a), marked P1 and P2.
	
The TD maps show a main intensity enhancement at 32 minutes (22:32 UT) and a series of additional enhancements before and after this peak, with an average temporal separation of $\delta t_\mathrm{EUV}$=1.7$\pm$0.2 min. These are labelled E1-E6 in panels a) and c) of Fig. \ref{td_maps}. At low distances along the slit there are a series of more localised guided enhancements which are visible in both TD maps, these are not sufficiently resolved due to their apparent period of 24 seconds, twice the time resolution of the data. These features are highlighted in panel b) of Fig. \ref{td_maps} by box B1. 

The gradient of a slope fitted to the main wave fronts can give a measurement of the propagation speed, however the scales involved are such that almost vertical fronts are obtained, not allowing an accurate estimate. However some propagating features are seen,  highlighted in Fig. \ref{td_maps} by the diagonal green lines, giving phase speeds of $\sim$ 1200 km s$^{-1}$, therefore we define the apparent phase speed to be $\geq$ 1200  km s$^{-1}$, which is consistent with the estimate made from Fig. \ref{prop_fig}.

To view the wave train front more clearly a time series was extracted from the running difference TD map at the distance marked by D in panel b) in Fig. \ref{td_maps}. This is plotted in panel c), with E1-E5 indicating the peaks of interest (E6 is missed as it is not prominent at the chosen distance along the slit).  The peaks are seen to be amplitude modulated. A Morlet wavelet spectra of the intensity time series is shown in the bottom panel of Fig. \ref{td_maps}. The solid white line corresponds to the 90 \% significance level \citep{1998BAMS...79...61T}. A powerful signal with a period of just below 2 minutes is present between 22:28 and 22:36 UT, reflecting the behaviour seen in the TD maps and time series, and agreeing with the period obtained. The 1 minute periodicity could be the second harmonic of the main, 2 minute signal, connected with nonlinear effects, as it appears when the amplitude of the main signal is higher.
	
There is a region of enhanced emission seen in 171 $\AA$ and 131 $\AA$, indicated in all three panels of Fig. \ref{3_wt} by the blue box. This lasts for the same duration as the series of enhancements and may be linked to the driving of the periodic wave train. In panel c) of Fig. \ref{3_wt} a propagating outflow is indicated, which has a different direction of propagation to the enhancements seen in 171 $\AA$, and is not periodic. This may be a jet of hot plasma related to the reconnection process indicated in Fig. \ref{3_wt} by the blue box. The direction of propagation does overlap with the path of the 171 $\AA$ enhancements, and may contribute to the complexity of the data in Fig. \ref{td_maps}.

\subsection{Radio observations}

\begin{figure}
\resizebox{\hsize}{!}{\includegraphics{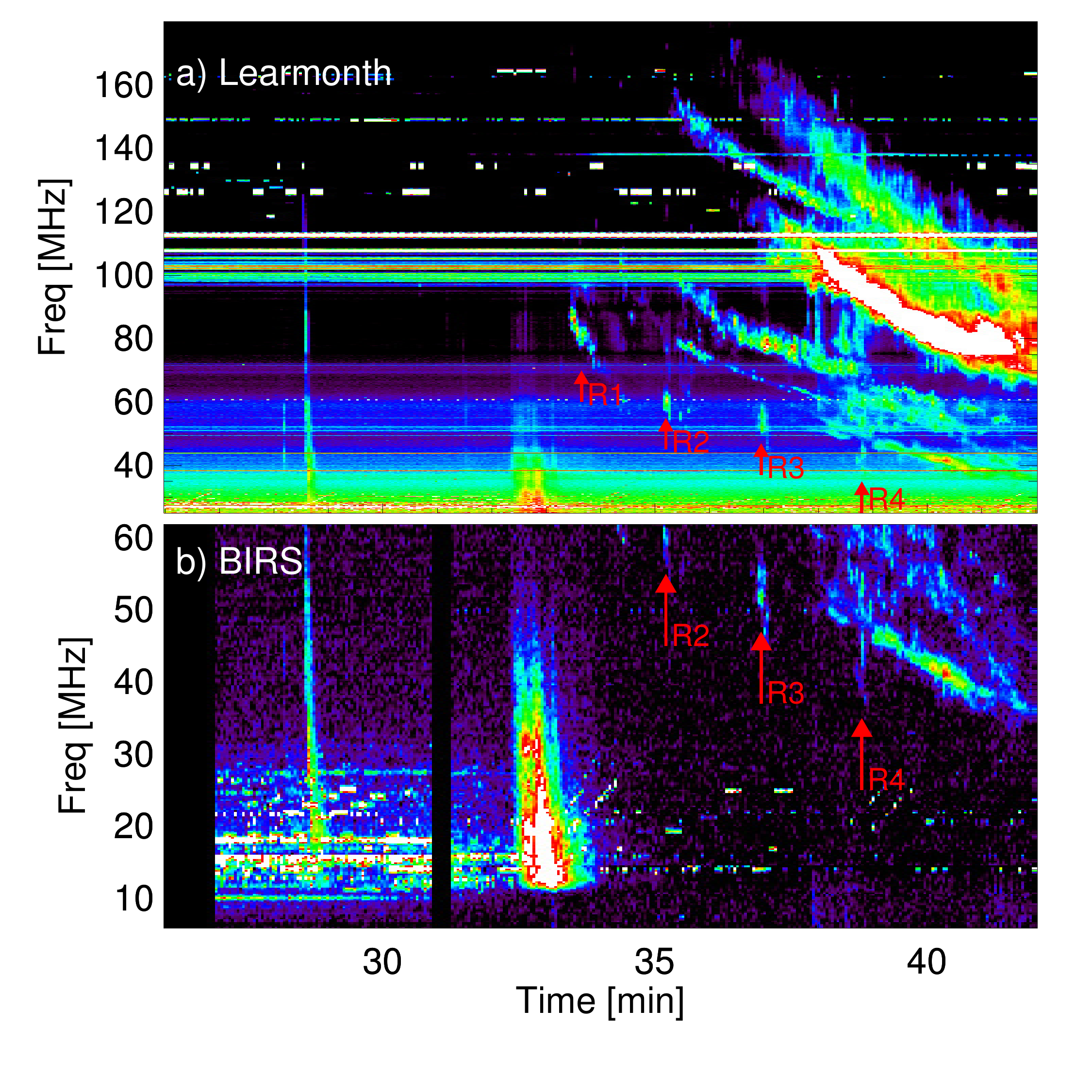}}
\caption{Learmonth, a), and BIRS, b), radio spectra in the ranges 25-170 MHz and 5-60 MHz respectively. Four regions of enhanced emission are indicated in the in a) by R1, R2, R3 and R4. R2-R4 are also indicated in b). The time axis refers to the time elapsed since 22:00 UT.}
\label{radio}
\end{figure}

\begin{figure}
\resizebox{\hsize}{!}{\includegraphics{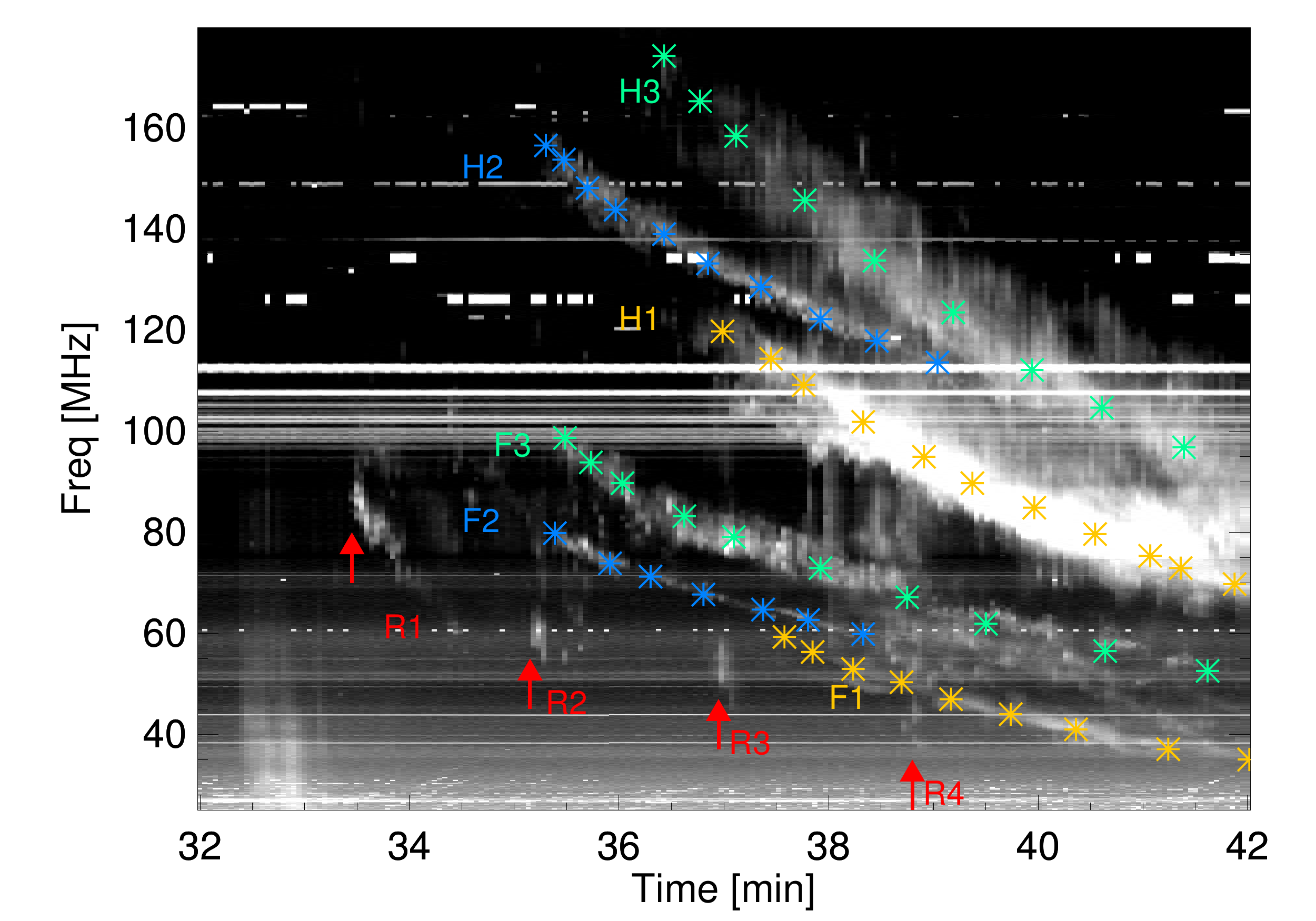}}
\caption{The Learmonth radio spectra. 3 lanes of fundamental emission are indicated by F1, F2 and F3. 3 lanes of harmonic emission are indicated by H1, H2 and H3.
Four discrete regions of enhanced emission are indicated by R1, R2, R3 and R4. The time axis refers to the time elapsed since 22:00 UT. }
\label{radio_zoom}
\end{figure}

\begin{figure*}
\resizebox{\hsize}{!}{\includegraphics{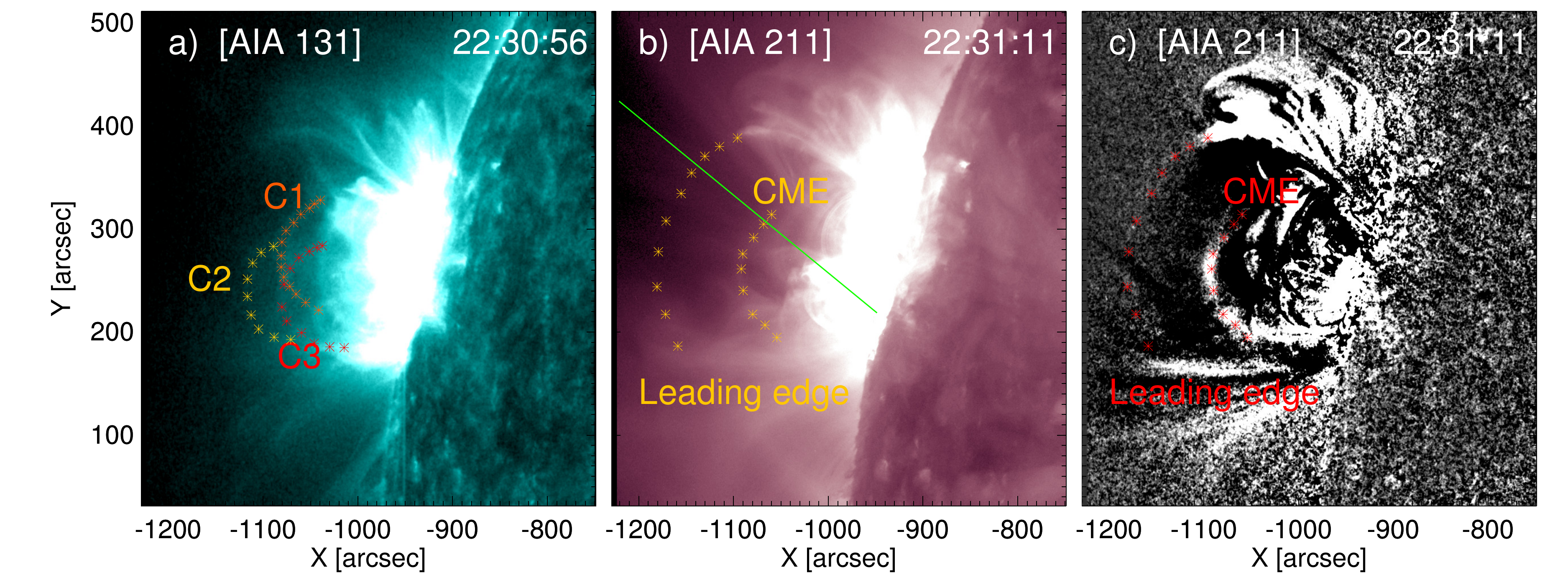}}
\caption{ Panel a): a 131 $\AA$ image showing three separate components of the expanding CME labelled as C1, C2 and C3. Panel b) and c): 211 $\AA$ images with the CME and the leading edge which precedes it indicated in orange in panel b) and red in the difference image in panel c). The solid green line in panel b) indicates the position of the slit used to analyse the expanding feature ahead of the CME. }
\label{3_cme}
\end{figure*}

\begin{figure}
\resizebox{\hsize}{!}{\includegraphics{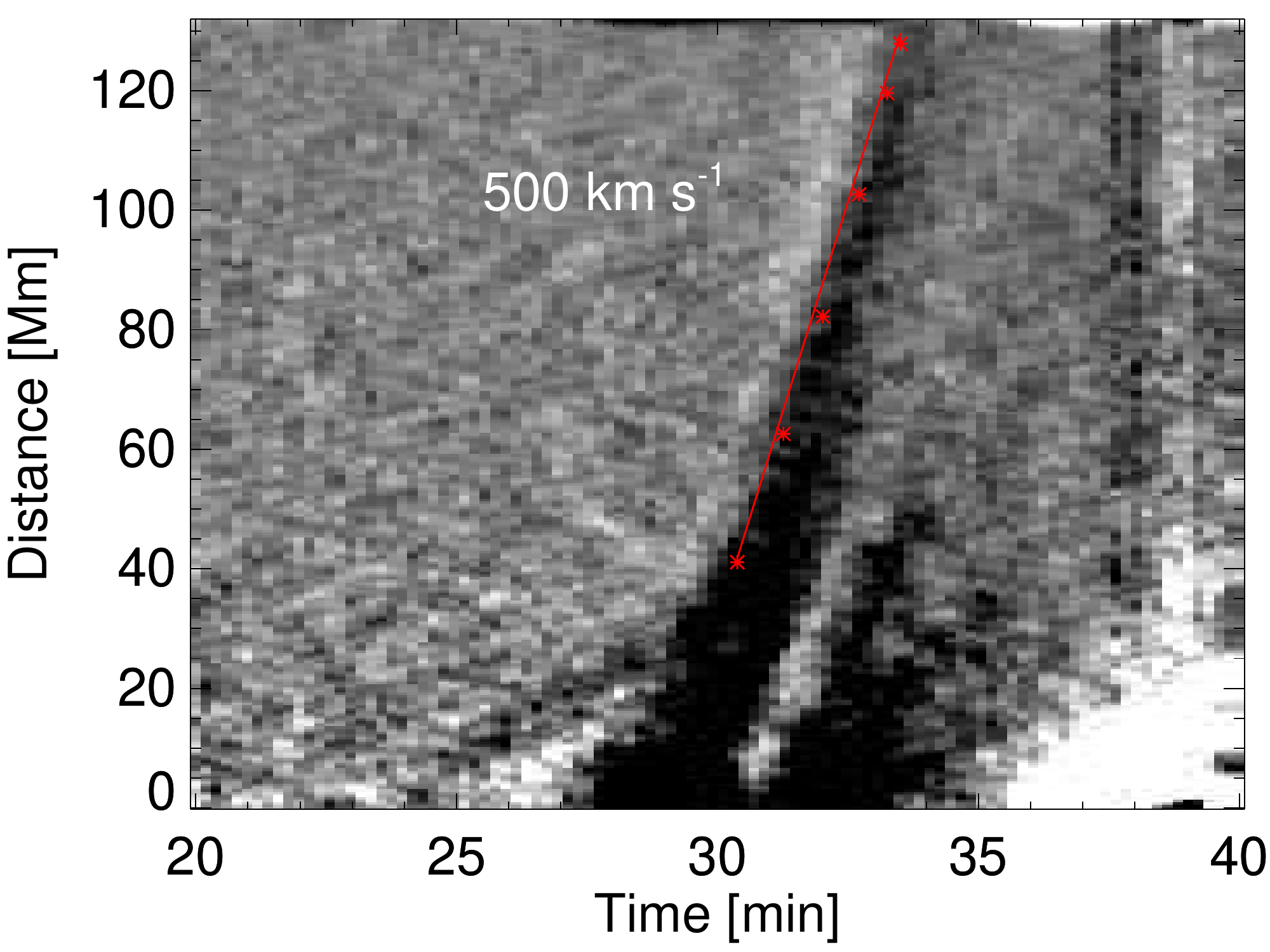}}
\caption{ A TD map from the 211 $\AA$ running difference images, formed from the slit marked in panel b) of Fig. \ref{3_cme}. The red points and linear fit mark the propagating feature, corresponding to a speed of 500 km s$^{-1}$. The time axis of the four panels refers to the time elapsed since 22:00 UT.}
\label{211_td}
\end{figure}

The dynamic radio spectra show four discrete narrowband short-lived features in a train at frequencies, and therefore densities, similar to the Type II burst. These features are labelled R1-R4 in Fig. \ref{radio}. R1 shows some Type-II-like drift, but in general the properties of these features do not match any of the classical solar radio burst types and therefore we will refer to them here as radio sparks for convenience. R1 is missing from the BIRS spectrograph as it lies outside the observational band, therefore we use the Learmonth data for the following analysis. 

The periodic sparks are centred on the following frequencies $F$=[83, 59, 54, 42] MHz. Using the equation for the plasma frequency given in the Introduction, and assuming that this emission is at the fundamental plasma frequency, these frequencies correspond to densities of $n_\mathrm{e}$=[8.47, 4.34, 3.56, 2.12] $\times 10^{7} \mathrm{cm}^{-3}$. Using the empirical formula for the height of these densities by rearranging the Newkirk formula given in the introduction yields heights above the base of the corona  of $Z$=[209, 288, 334, 418] Mm. The periodicity of the radio bursts is: $P_\mathrm{r}$=1.78$\pm$0.04 min measured from the beginning of each spark.

A type II burst also observed in this event includes three separate strong lanes of emission, labelled as H1, H2 and H3 in Fig. \ref{radio_zoom}. Using the plasma frequency and the Newkirk model as described above, the three lanes give us speeds of 630, 380, 550 km s$^{-1}$, respectively. These are interpreted as the harmonic emission from the three weaker fundamental emission lanes, labelled as F1, F2 and F3. This indicates the periodic sparks R1-R4 are not a typical type II emission lane as they do not have a stronger harmonic component. 

The time between the periodic sparks and the relative change in height from the calculated density gives an estimate of the emission location speed of $v_\mathrm{emn}$= 630 km s$^{-1}$. 

\subsection{Further analysis}

We analyse the association between the periodic radio sparks and the CME to make inferences about the location at which the radio sparks are produced. The height of the CME leading edge indicated in panel b) and c) of Fig. \ref{3_cme} is 247 Mm at the time of the first radio burst. A TD map, shown in Fig. \ref{211_td}, taken from the slit indicated in panel b) of Fig. \ref{3_cme} gives a speed of $\sim$ 500 km s$^{-1}$ for the CME leading edge. This slit position was chosen as it offers the best signal to noise ratio for detecting the propagation. Extrapolating this forward to the times of the subsequent radio sparks gives heights of $Z$=[247, 298, 349, 400] Mm, which are roughly consistent with the emission heights of the radio sparks derived above. 

From the upper and lower frequencies of each radio spark it is possible to obtain an estimate for the vertical extent of the emission region. From the frequencies the corresponding densities are obtained assuming the emission is at the electron plasma frequency, which are then used to calculate the upper and lower heights of the emission from the Newkirk formula. The resulting vertical lengths of the emission region for each spark are $L$=[29, 24, 26, 32] Mm. 

The complex nature of the expanding CME structure is highlighted in panel a) of Fig. \ref{3_cme}. Three separate expanding structures were identified from the series of images and are labelled as C1, C2 and C3. This series of expanding structures, and the complex geometry of the active region, provide adequate mechanisms to produce the 3 fundamental and harmonic emission lanes highlighted in Fig. \ref{radio_zoom}. This is supported by the speeds derived from the drifts in the radio spectrum, which are in the range of typical CME velocities in the low corona.
 
Finally, if we assume the wave train fronts cause the radio sparks when they reach the propagating feature ahead of the CME indicated as \lq leading edge\rq\ in Fig. \ref{3_cme}, then the temporal separation ($t_\mathrm{r}$-$t_\mathrm{EUV}$) and height of the emission, Z or $Z_\mathrm{CME}$, can be used to estimate the average wave train propagation speed between the active region and the emission location. Matching the first EUV enhancement (E1) with the first radio burst (R1) gives $v_\mathrm{EUV}$= [700, 970, 1120, 1390] km s$^{-1}$. Matching the strongest wave train enhancement (E3) with R1 gives $v_\mathrm{EUV} >$ 4000 km s$^{-1}$, which is unrealistic for fast magnetoacoustic waves in the corona. The first set of speeds are approximately consistent with the speeds measured lower in the corona for the propagating wave train fronts.


\section{Discussion and Conclusions}

\begin{figure*}
\includegraphics[trim={10cm 20cm 25cm 10cm}, clip, width=\textwidth , keepaspectratio]{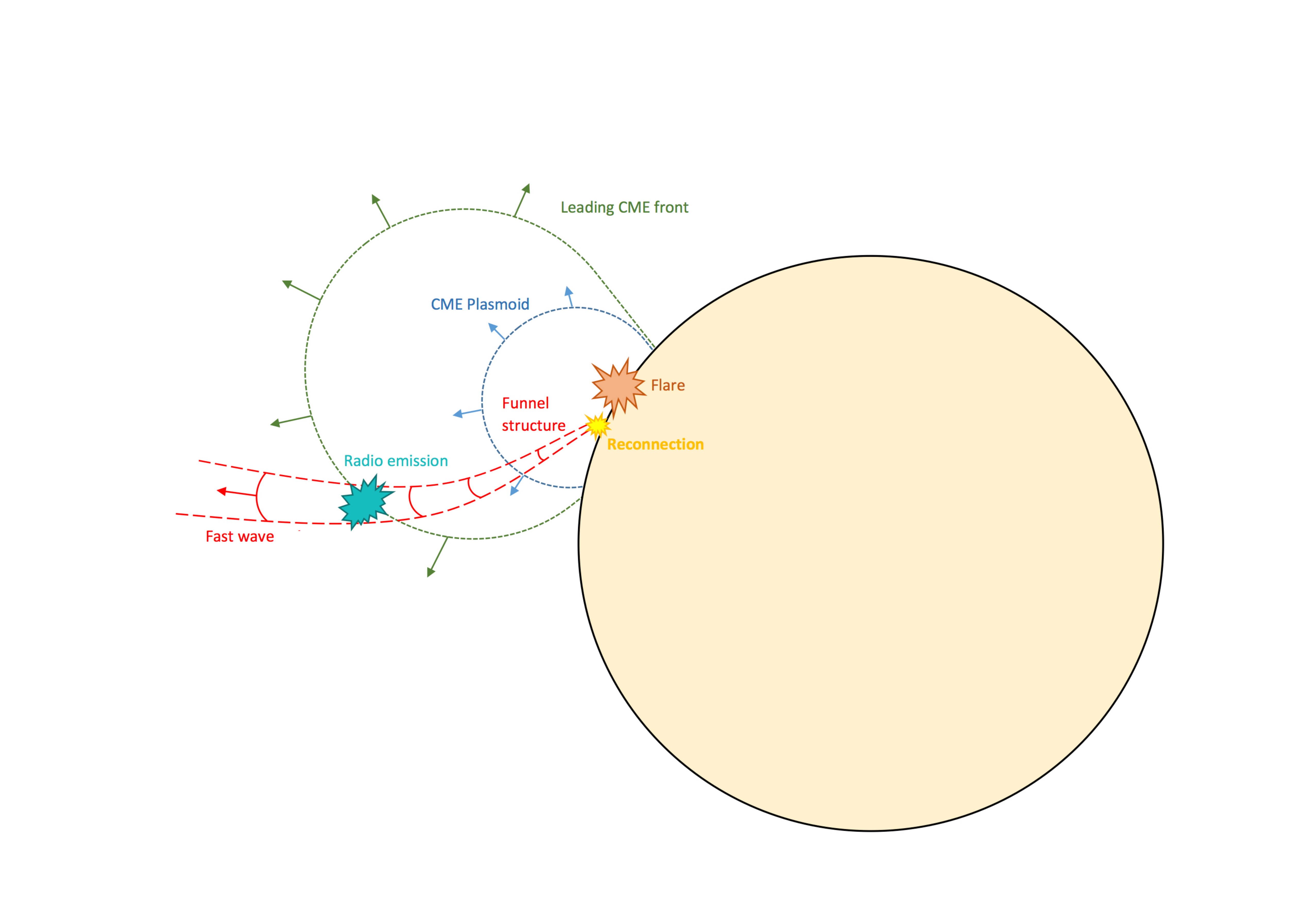}
\caption{A schematic synopsis of the event. A flare occurs which is followed by a CME comprising of the leading edge or EUV wave (green) and the main CME plasmoid (blue). A funnel structure (red) within the active region is seen to host a series of rapidly propagating quasi-periodic waves or jets. A brightening is observed at the base of this structure and is interpreted as a reconnection site. After a certain delay periodic radio \lq sparks\rq are observed, which occur at an estimated height consistent with the leading feature of the CME, and a periodicity consistent with the fast wave period.}
\label{sketch}
\end{figure*}

Quasi-periodic EUV intensity disturbances are found to propagate along a guiding funnel structure during a flaring event, beginning at 22:27:59 UT. The period is $P_\mathrm{EUV}$=1.7 $\pm$ 0.2 min, which becomes more pronounced with distance along the wave guide. The CME plasmoid associated with the flaring event is seen to interact with the active region at 22:27:56 UT, resulting in a region of enhanced emission seen in all channels analysed near the base of the funnel structure. Thus, we can assume that the periodic wave train is induced by the CME interaction with background structures, possibly due to the resulting reconnection. A series of small radio bursts, or sparks, occurs during the CME expansion prior to the type II emission. These have a period of $P_\mathrm{r}$=1.78$\pm$0.04 min, making it a reasonable assumption that they are linked to the periodic fast wave train. 

The 171 $\AA$ intensity disturbances can be interpreted as a series of guided fast magnetoacoustic waves \citep[see][]{2011ApJ...740L..33O, 2013A&A...560A..97P}, which may be formed by the dispersive evolution of a pulse excited in the guiding structure, or by a periodic quasi-harmonic driver. Some wave train fronts are clearly seen to propagate with projected speeds of 1200 km s$^{-1}$, which are consistent with previous fast wave train observations. Some previous observations have interpreted the observed wave trains as the result of repetitive magnetic reconnection associated with the flare, or another mechanism which periodically excites broadband pulses of fast waves. Recent modelling results from \cite{2015ApJ...800..111Y} have confirmed this as a viable mechanism for the production of a series of fast waves with phase speeds and observational signatures which match observations.

Other possibilities for the nature of the enhancements exist, such as periodic jets. However jets are normally multi-thermal and are therefore seen in multiple channels \citep{2009SoPh..259...87N}. In our observations no signature of the intensity enhancements was seen at other AIA wavelengths. Jets are also normally seen as narrow structures, which are more long lived than our observations, and do not have low periodicities of several minutes. It is possible that a superposition of fast waves and ejections is present, which would explain the complex dynamics observed (see animation provided on-line).

The triple band type II burst is clearly resolved in the radio spectra, and it is possible to match the strong harmonic emission with their fundamental counterparts.  The series of periodic radio sparks which precede these do not correspond to any of the observed lanes, leading us to interpret them as a separate phenomenon, which causes emission at the local plasma frequency without a second harmonic component. They approximately follow the same drifting trend as the fundamental components however, meaning their emission location may exhibit the same dynamic behaviour as the CME which produces the shocks. We have found drifting velocities of H1, H2 and H3 to be 630, 380 and 550 km s$^{-1}$, leading us to interpret them as emission associated with spatially separated shock waves driven by different parts of the expanding CME.

Since the rapidly-propagating periodic wave train observed in the low corona and the series of radio sparks have almost equal periodicities we consider the periodic waves seen in the EUV band to be the drivers of the periodic radio sparks. The slight offset in the detected periods may support this, as we expect the time delay between the radio sparks to be longer than the wave train period, due to the propagation of the CME leading edge. As discussed above, this allows us to make an estimate of the transit velocity from the base of the guiding structure to the radio burst emission location using the inferred distance and time delay between the observations. The most reasonable estimates came from matching wave train fronts E1 or E2 with R1, which gives transit velocities in the range 800 - 2000 km s$^{-1}$, which are roughly consistent with the velocities measured from the TD map in Fig. \ref{td_maps}, and are in the range of fast waves. There is a large degree of uncertainty associated with these estimates, however it is clear that the time delay between the EUV and radio features is large enough to exclude energetic particles accelerated in the active region as the driver of emission at these heights, due to their characteristic high propagation velocities. 

Our proposed scenario is similar to the \lq cannibalism\rq of CMEs, when one faster CME (ejected later) catches another slower CME (ejected earlier). In some of these cases it has been found that the related type II radio burst emission can be enhanced during the process of two CMEs merging  \citep[e.g][]{2001ApJ...548L..91G, 2012ApJ...748...66M}.

The different spectral aspects of the radio sparks can be explained by the properties of the wave fronts, which may have different temporal and spatial extents, similar to the snapshots of the fast magnetoacoustic waves generated by geometrical dispersion in a plasma funnel in \cite{2013A&A...560A..97P}. The first spark exhibits a frequency drift, which can be explained by a broad wave front, such as the one indicated in Fig. \ref{3_wt}. Additionally, from Fig. \ref{td_maps} panel a) it can be seen that E1 is more temporally broad than the following peaks, which could give rise to the drift seen in R1. From the frequency range each spark covers the vertical extent of the emission region for each was calculated, giving L=[29, 24, 26, 32] Mm for R1 - R4. These values support our interpretation that the emission is generated in a localised region corresponding to a feature of finite vertical width, such as the expanding front ahead of the CME.

The estimated heights at which the radio sparks (R1-R4) are generated roughly match the positions of the CME leading edge marked in Fig. \ref{3_cme}.  Additionally, the trend of the sparks in the radio spectra matches the drifting of the fundamental emission lanes, indicating a definite link to the kinematics of the CME. The CME leading edge may be a developing EUV wave before it has decoupled from the expanding CME.

Possible excitation scenarios to produce the periodic radio emission include: steepening of the periodic wave train fronts which shock in the medium of the expanding CME leading edge, emission due to the compression of the medium between the CME leading edge and the approaching fast wave train, or alternative emission mechanisms such as the cyclotron-maser mechanism discussed in \cite{2005ApJ...621.1129W}. These scenarios would produce accelerated electrons, the bump-on-tail instability, and subsequent emission of the radio waves with the frequency corresponding to the local electron plasma frequency.

Another possibility exists to explain the periodic radio emission without the inclusion of the CME features. Panel c) of Fig. \ref{td_maps} shows the series of EUV enhancements vary in amplitude. Fast wave steeping depends on the waves amplitude, so different cycles of oscillation in the amplitude-modulated dispersively-formed fast wave train \citep[e.g.][]{2012A&A...546A..49J, 2013A&A...560A..97P, 2014ApJ...788...44M} will shock at different heights in the corona, which could produce the drifting of the sparks from high to low frequency as observed. However, this scenario does not explain why the appearance of the sparks in the radio spectrum matches the drift of the type II bursts.

\section{Summary}
We analysed a flaring event and the associated CME and periodic waves with SDO data, and the corresponding radio features with Learmonth and BIRS data. A series of finite-bandwidth, short-duration isolated radio features drifting from high to low frequency are observed. The period of these radio sparks, $P_\mathrm{r}$=1.78$\pm$0.04 min, matches the period of the rapidly propagating wave train observed at 171 $\AA$, $P_\mathrm{EUV}$=1.7 $\pm$ 0.2 min. The speed of the radio emission location, 630 km s$^{-1}$, estimated from the instant frequencies of the radio sparks, is of the same order as the speed of the CME and its leading edge, 500 km s$^{-1}$. The calculated height of the radio emission matches the observed (and then projected forward using the observed velocity) location of the leading edge of the CME. Using the time delay between the wave train fronts and radio sparks and the height of the emission, propagation speeds in the range of fast magnetoacoustic waves are obtained.

We interpret the entire observations with the following physical scenario. A series of fast waves are produced by the active region during a flare, during an observed magnetic reconnection event. The waves propagate upwards along a funnel plasma structure, and interact with the CME leading edge, or some associated disturbance that propagates slower than the fast wave train. This results in the acceleration of electrons, the bump-on-tail instability, and emission of radio waves with the frequency corresponding to the local electron plasma frequency, appearing as quasi-periodic sparks in the radio spectrograph (see Fig. \ref{sketch}). Theoretical modelling of the potential emission mechanisms is needed.

\begin{acknowledgements}
The work was supported by the European Research Council under the SeismoSun Research Project No. 321141 (CRG, VMN), and the STFC consolidated grant ST/L000733/1 (GN, VMN). IVZ is supported by the RFBR grant No. 15-32-21078. The data is used courtesy of the SDO/AIA team. We thank the referee for their helpful and constructive comments. 
\end{acknowledgements}

\bibliographystyle{aa} 
\bibliography{radio} 

\begin{thebibliography}{49}
\expandafter\ifx\csname natexlab\endcsname\relax\def\natexlab#1{#1}\fi

\bibitem[{{Anfinogentov} {et~al.}(2015){Anfinogentov}, {Nakariakov}, \&
  {Nistic{\`o}}}]{2015A&A...583A.136A}
{Anfinogentov}, S.~A., {Nakariakov}, V.~M., \& {Nistic{\`o}}, G. 2015, \aap,
  583, A136

\bibitem[{{Chen} {et~al.}(2005){Chen}, {Fang}, \&
  {Shibata}}]{2005ApJ...622.1202C}
{Chen}, P.~F., {Fang}, C., \& {Shibata}, K. 2005, \apj, 622, 1202

\bibitem[{{De Moortel}(2009)}]{2009SSRv..149...65D}
{De Moortel}, I. 2009, \ssr, 149, 65

\bibitem[{{De Moortel} \& {Nakariakov}(2012)}]{2012RSPTA.370.3193D}
{De Moortel}, I. \& {Nakariakov}, V.~M. 2012, Royal Society of London
  Philosophical Transactions Series A, 370, 3193

\bibitem[{{Erickson}(1997)}]{1997PASA...14..278E}
{Erickson}, W.~C. 1997, \pasa, 14, 278

\bibitem[{{Gallagher} \& {Long}(2011)}]{2011SSRv..158..365G}
{Gallagher}, P.~T. \& {Long}, D.~M. 2011, \ssr, 158, 365

\bibitem[{{Goddard} {et~al.}(2016){Goddard}, {Nistic{\`o}}, {Nakariakov}, \&
  {Zimovets}}]{2016A&A...585A.137G}
{Goddard}, C.~R., {Nistic{\`o}}, G., {Nakariakov}, V.~M., \& {Zimovets}, I.~V.
  2016, \aap, 585, A137

\bibitem[{{Gopalswamy} {et~al.}(2001){Gopalswamy}, {Yashiro}, {Kaiser},
  {Howard}, \& {Bougeret}}]{2001ApJ...548L..91G}
{Gopalswamy}, N., {Yashiro}, S., {Kaiser}, M.~L., {Howard}, R.~A., \&
  {Bougeret}, J.-L. 2001, \apjl, 548, L91

\bibitem[{{Jel{\'{\i}}nek} {et~al.}(2012){Jel{\'{\i}}nek}, {Karlick{\'y}}, \&
  {Murawski}}]{2012A&A...546A..49J}
{Jel{\'{\i}}nek}, P., {Karlick{\'y}}, M., \& {Murawski}, K. 2012, \aap, 546,
  A49

\bibitem[{Kennewell \& Steward(2003)}]{learmonth...2003}
Kennewell, J. \& Steward, G. 2003, Solar Radio Spectrograph [SRS] Data Viewer
  [Srsdisplay] (Sydney: IPS Radio and Space Serv.)

\bibitem[{{Kupriyanova} {et~al.}(2010){Kupriyanova}, {Melnikov}, {Nakariakov},
  \& {Shibasaki}}]{2010SoPh..267..329K}
{Kupriyanova}, E.~G., {Melnikov}, V.~F., {Nakariakov}, V.~M., \& {Shibasaki},
  K. 2010, \solphys, 267, 329

\bibitem[{{Lemen} {et~al.}(2012){Lemen}, {Title}, {Akin}, {Boerner}, {Chou},
  {Drake}, {Duncan}, {Edwards}, {Friedlaender}, {Heyman}, {Hurlburt}, {Katz},
  {Kushner}, {Levay}, {Lindgren}, {Mathur}, {McFeaters}, {Mitchell}, {Rehse},
  {Schrijver}, {Springer}, {Stern}, {Tarbell}, {Wuelser}, {Wolfson}, {Yanari},
  {Bookbinder}, {Cheimets}, {Caldwell}, {Deluca}, {Gates}, {Golub}, {Park},
  {Podgorski}, {Bush}, {Scherrer}, {Gummin}, {Smith}, {Auker}, {Jerram},
  {Pool}, {Soufli}, {Windt}, {Beardsley}, {Clapp}, {Lang}, \&
  {Waltham}}]{2012SoPh..275...17L}
{Lemen}, J.~R., {Title}, A.~M., {Akin}, D.~J., {et~al.} 2012, \solphys, 275, 17

\bibitem[{{Liu} {et~al.}(2012){Liu}, {Ofman}, {Nitta}, {Aschwanden},
  {Schrijver}, {Title}, \& {Tarbell}}]{2012ApJ...753...52L}
{Liu}, W., {Ofman}, L., {Nitta}, N.~V., {et~al.} 2012, \apj, 753, 52

\bibitem[{{Liu} {et~al.}(2011){Liu}, {Title}, {Zhao}, {Ofman}, {Schrijver},
  {Aschwanden}, {De Pontieu}, \& {Tarbell}}]{2011ApJ...736L..13L}
{Liu}, W., {Title}, A.~M., {Zhao}, J., {et~al.} 2011, \apjl, 736, L13

\bibitem[{{Magdaleni{\'c}} {et~al.}(2005){Magdaleni{\'c}}, {Vr{\v s}nak},
  {Zlobec}, {Messerotti}, \& {Temmer}}]{2005ASSL..320..259M}
{Magdaleni{\'c}}, J., {Vr{\v s}nak}, B., {Zlobec}, P., {Messerotti}, M., \&
  {Temmer}, M. 2005, in Astrophysics and Space Science Library, Vol. 320, Solar
  Magnetic Phenomena, ed. A.~{Hanslmeier}, A.~{Veronig}, \& M.~{Messerotti},
  259--262

\bibitem[{{Mann} {et~al.}(1995){Mann}, {Classen}, \&
  {Aurass}}]{1995A&A...295..775M}
{Mann}, G., {Classen}, T., \& {Aurass}, H. 1995, \aap, 295, 775

\bibitem[{{Mart{\'{\i}}nez Oliveros} {et~al.}(2012){Mart{\'{\i}}nez Oliveros},
  {Raftery}, {Bain}, {Liu}, {Krupar}, {Bale}, \&
  {Krucker}}]{2012ApJ...748...66M}
{Mart{\'{\i}}nez Oliveros}, J.~C., {Raftery}, C.~L., {Bain}, H.~M., {et~al.}
  2012, \apj, 748, 66

\bibitem[{{McLean}(1967)}]{1967PASAu...1...47M}
{McLean}, D.~J. 1967, Proceedings of the Astronomical Society of Australia, 1,
  47

\bibitem[{{M{\'e}sz{\'a}rosov{\'a}} {et~al.}(2014){M{\'e}sz{\'a}rosov{\'a}},
  {Karlick{\'y}}, {Jel{\'{\i}}nek}, \& {Ryb{\'a}k}}]{2014ApJ...788...44M}
{M{\'e}sz{\'a}rosov{\'a}}, H., {Karlick{\'y}}, M., {Jel{\'{\i}}nek}, P., \&
  {Ryb{\'a}k}, J. 2014, \apj, 788, 44

\bibitem[{{M{\'e}sz{\'a}rosov{\'a}} {et~al.}(2011){M{\'e}sz{\'a}rosov{\'a}},
  {Karlick{\'y}}, \& {Ryb{\'a}k}}]{2011SoPh..273..393M}
{M{\'e}sz{\'a}rosov{\'a}}, H., {Karlick{\'y}}, M., \& {Ryb{\'a}k}, J. 2011,
  \solphys, 273, 393

\bibitem[{{M{\'e}sz{\'a}rosov{\'a}} {et~al.}(2009){M{\'e}sz{\'a}rosov{\'a}},
  {Karlick{\'y}}, {Ryb{\'a}k}, \& {Ji{\v r}i{\v c}ka}}]{2009ApJ...697L.108M}
{M{\'e}sz{\'a}rosov{\'a}}, H., {Karlick{\'y}}, M., {Ryb{\'a}k}, J., \& {Ji{\v
  r}i{\v c}ka}, K. 2009, \apjl, 697, L108

\bibitem[{{Moreton}(1960)}]{1960AJ.....65U.494M}
{Moreton}, G.~E. 1960, \aj, 65, 494

\bibitem[{{Nakariakov} \& {Melnikov}(2009)}]{2009SSRv..149..119N}
{Nakariakov}, V.~M. \& {Melnikov}, V.~F. 2009, \ssr, 149, 119

\bibitem[{Nakariakov {et~al.}(2016)Nakariakov, Pilipenko, Heilig, Jel{\'i}nek,
  Karlick{\'y}, Klimushkin, Kolotkov, Lee, Nistic{\`o}, Doorsselaere, Verth, \&
  Zimovets}]{Nakariakov2016}
Nakariakov, V.~M., Pilipenko, V., Heilig, B., {et~al.} 2016, Space Science
  Reviews, 1

\bibitem[{{Nakariakov} \& {Verwichte}(2005)}]{2005LRSP....2....3N}
{Nakariakov}, V.~M. \& {Verwichte}, E. 2005, Living Reviews in Solar Physics,
  2, 3

\bibitem[{{Newkirk}(1961)}]{1961ApJ...133..983N}
{Newkirk}, Jr., G. 1961, \apj, 133, 983

\bibitem[{{Nistic{\`o}} {et~al.}(2009){Nistic{\`o}}, {Bothmer}, {Patsourakos},
  \& {Zimbardo}}]{2009SoPh..259...87N}
{Nistic{\`o}}, G., {Bothmer}, V., {Patsourakos}, S., \& {Zimbardo}, G. 2009,
  \solphys, 259, 87

\bibitem[{{Nistic{\`o}} {et~al.}(2014){Nistic{\`o}}, {Pascoe}, \&
  {Nakariakov}}]{2014A&A...569A..12N}
{Nistic{\`o}}, G., {Pascoe}, D.~J., \& {Nakariakov}, V.~M. 2014, \aap, 569, A12

\bibitem[{{Ofman} {et~al.}(2011){Ofman}, {Liu}, {Title}, \&
  {Aschwanden}}]{2011ApJ...740L..33O}
{Ofman}, L., {Liu}, W., {Title}, A., \& {Aschwanden}, M. 2011, \apjl, 740, L33

\bibitem[{{Parks} \& {Winckler}(1969)}]{1969ApJ...155L.117P}
{Parks}, G.~K. \& {Winckler}, J.~R. 1969, \apjl, 155, L117

\bibitem[{{Pascoe} {et~al.}(2013){Pascoe}, {Nakariakov}, \&
  {Kupriyanova}}]{2013A&A...560A..97P}
{Pascoe}, D.~J., {Nakariakov}, V.~M., \& {Kupriyanova}, E.~G. 2013, \aap, 560,
  A97

\bibitem[{{Patsourakos} {et~al.}(2009){Patsourakos}, {Vourlidas}, {Wang},
  {Stenborg}, \& {Thernisien}}]{2009SoPh..259...49P}
{Patsourakos}, S., {Vourlidas}, A., {Wang}, Y.~M., {Stenborg}, G., \&
  {Thernisien}, A. 2009, \solphys, 259, 49

\bibitem[{{Pick} \& {Vilmer}(2008)}]{2008A&ARv..16....1P}
{Pick}, M. \& {Vilmer}, N. 2008, \aapr, 16, 1

\bibitem[{{Pugh} {et~al.}(2016){Pugh}, {Armstrong}, {Nakariakov}, \&
  {Broomhall}}]{2016MNRAS.459.3659P}
{Pugh}, C.~E., {Armstrong}, D.~J., {Nakariakov}, V.~M., \& {Broomhall}, A.-M.
  2016, \mnras, 459, 3659

\bibitem[{{Robbrecht} {et~al.}(2009){Robbrecht}, {Berghmans}, \& {Van der
  Linden}}]{2009ApJ...691.1222R}
{Robbrecht}, E., {Berghmans}, D., \& {Van der Linden}, R.~A.~M. 2009, \apj,
  691, 1222

\bibitem[{{Schmidt} \& {Cairns}(2012)}]{2012JGRA..117.4106S}
{Schmidt}, J.~M. \& {Cairns}, I.~H. 2012, Journal of Geophysical Research
  (Space Physics), 117, A04106

\bibitem[{{Shen} \& {Liu}(2012)}]{2012ApJ...753...53S}
{Shen}, Y. \& {Liu}, Y. 2012, \apj, 753, 53

\bibitem[{{Smerd} {et~al.}(1974){Smerd}, {Sheridan}, \&
  {Stewart}}]{1974IAUS...57..389S}
{Smerd}, S.~F., {Sheridan}, K.~V., \& {Stewart}, R.~T. 1974, in IAU Symposium,
  Vol.~57, Coronal Disturbances, ed. G.~A. {Newkirk}, 389

\bibitem[{{Sych} {et~al.}(2009){Sych}, {Nakariakov}, {Karlicky}, \&
  {Anfinogentov}}]{2009A&A...505..791S}
{Sych}, R., {Nakariakov}, V.~M., {Karlicky}, M., \& {Anfinogentov}, S. 2009,
  \aap, 505, 791

\bibitem[{{Torrence} \& {Compo}(1998)}]{1998BAMS...79...61T}
{Torrence}, C. \& {Compo}, G.~P. 1998, Bulletin of the American Meteorological
  Society, 79, 61

\bibitem[{{Wang}(2011)}]{2011SSRv..158..397W}
{Wang}, T. 2011, \ssr, 158, 397

\bibitem[{{Warmuth}(2015)}]{2015LRSP...12....3W}
{Warmuth}, A. 2015, Living Reviews in Solar Physics, 12

\bibitem[{{Wu} {et~al.}(2005){Wu}, {Wang}, {Zhou}, {Wang}, \&
  {Yoon}}]{2005ApJ...621.1129W}
{Wu}, C.~S., {Wang}, C.~B., {Zhou}, G.~C., {Wang}, S., \& {Yoon}, P.~H. 2005,
  \apj, 621, 1129

\bibitem[{{Yang} {et~al.}(2015){Yang}, {Zhang}, {He}, {Peter}, {Tu}, {Wang},
  {Zhang}, \& {Feng}}]{2015ApJ...800..111Y}
{Yang}, L., {Zhang}, L., {He}, J., {et~al.} 2015, \apj, 800, 111

\bibitem[{{Yu} {et~al.}(2013){Yu}, {Nakariakov}, {Selzer}, {Tan}, \&
  {Yan}}]{2013ApJ...777..159Y}
{Yu}, S., {Nakariakov}, V.~M., {Selzer}, L.~A., {Tan}, B., \& {Yan}, Y. 2013,
  \apj, 777, 159

\bibitem[{{Yuan} {et~al.}(2013){Yuan}, {Shen}, {Liu}, {Nakariakov}, {Tan}, \&
  {Huang}}]{2013A&A...554A.144Y}
{Yuan}, D., {Shen}, Y., {Liu}, Y., {et~al.} 2013, \aap, 554, A144

\bibitem[{{Zaitsev}(1966)}]{1966SvA.....9..572Z}
{Zaitsev}, V.~V. 1966, \sovast, 9, 572

\bibitem[{{Zimovets} {et~al.}(2012){Zimovets}, {Vilmer}, {Chian}, {Sharykin},
  \& {Struminsky}}]{2012A&A...547A...6Z}
{Zimovets}, I., {Vilmer}, N., {Chian}, A.~C.-L., {Sharykin}, I., \&
  {Struminsky}, A. 2012, \aap, 547, A6

\bibitem[{{Zimovets} \& {Sadykov}(2015)}]{2015AdSpR..56.2811Z}
{Zimovets}, I.~V. \& {Sadykov}, V.~M. 2015, Advances in Space Research, 56,
  2811

\end{thebibliography}


\end{document}